\documentclass[prd,
twocolumn,
superscriptaddress,nofootinbib,%
tightenlines,shownopacs]{revtex4}
\usepackage{eepic}
\usepackage{indentfirst}
\usepackage{mathrsfs}
\usepackage{fancyhdr}
\usepackage{ulem}
\usepackage{float}
\usepackage{graphicx}
\usepackage[
colorlinks=true, linkcolor=black, breaklinks=true, urlcolor=blue,
citecolor=green]{hyperref}
\usepackage{epstopdf}
\usepackage{bm,bbm}
\usepackage{amsmath}
\usepackage{amssymb}

\usepackage{tabularx}
\usepackage{subfigure}
\usepackage{epsfig}
\usepackage{color}
\usepackage{slashed}
\usepackage{hyperref}

\newcommand{\be}{\begin{equation}}
\newcommand{\ee}{\end{equation}}
\newcommand{\bea}{\begin{eqnarray}}
\newcommand{\eea}{\end{eqnarray}}

\allowdisplaybreaks

\begin{document}
\pagestyle{plain}

\title {\boldmath\Large A study of $\rho-\omega$ mixing in resonance chiral theory }
\author{ Yun-Hua~Chen}\thanks{yhchen@ustb.edu.cn}
\affiliation{School of Mathematics and Physics, \\University of
Science and Technology Beijing, Beijing 100083, China}
\author{ De-Liang Yao}\thanks{Corresponding author: deliang.yao@ific.uv.es}
\affiliation{Instituto de F\'{\i}sica Corpuscular (centro mixto
CSIC-UV), \\ Institutos de Investigaci\'on de Paterna, Apartado
22085, 46071, Valencia, Spain}
\author{ Han-Qing Zheng}\thanks{ zhenghq@pku.edu.cn }
\affiliation{Department of Physics and State Key Laboratory of
Nuclear Physics and Technology,\\ Peking University, Beijing 100871,
China} \affiliation{Collaborative Innovation Center of Quantum
Matter, Beijing 100871, China}

\begin{abstract}
The strong and electromagnetic corrections to $\rho-\omega$ mixing
are calculated using a SU(2) version of resonance chiral theory up
to next-to-leading orders in $1/N_C$ expansion, respectively.  Up to
our accuracy,  the effect of the momentum dependence of
$\rho-\omega$ mixing is incorporated due to the inclusion of loop
contributions. We analyze the impact of $\rho-\omega$ mixing on the
pion vector form factor by performing numerical fit to the data
extracted from $e^+e^-\rightarrow \pi^+\pi^-$ and $\tau\rightarrow
\nu_{\tau}2\pi$, while the decay width of $\omega\rightarrow
\pi^+\pi^-$ is taken into account as a constraint. It is found that
the momentum dependence is significant in a good description of the
experimental data. In addition, based on the fitted values of the
involved parameters, we analyze the decay width of $\omega
\rightarrow \pi^+\pi^-$, which turns out to be highly dominated by
the $\rho-\omega$ mixing effect.

\end{abstract}

\maketitle

\newpage

\section{Introduction}
The study of $\rho-\omega$ mixing is a very
interesting subject in hadron physics both theoretically and experimentally. The
inclusion of $\rho-\omega$ mixing effect is crucial for a good
description of the pion vector form factor in $e^+e^- \rightarrow
\pi^+\pi^-$ process, which quantifies the hadronic vacuum
polarization contribution to the anomalous magnetic moment of the
muon. On the experimental side, several experimental
collaborations, such as KLOE~\cite{KLOE2011,KLOE2013} and BESIII~\cite{BESIII}, have recently launched measurements of the
$e^+e^-\rightarrow\pi^+\pi^-$ with high statistics and high precision.

The $\rho-\omega$ mixing amplitude was assumed to be a constant or
momentum-independent in the early stage of previous studies~\cite{Glashow,Renard}. The
authors of Ref.~\cite{Goldman} suspect the validity of the constant
assumption and, based on a quark loop mechanism of $\rho-\omega$ mixing, they found that the mixing amplitude is
significantly momentum-dependent. Since then,  the use of various loop mechanisms
for $\rho-\omega$ mixing is triggered in different models such as extended
Nambu-Jona-Lasinio (NJL) model~\cite{Shakin}, the global color model
~\cite{Mitchell}, the hidden local symmetry
model~\cite{Benayoun00,Benayoun01,Benayoun08}, and the chiral
constituent quark model~\cite{MLYan98,MLYan00}.

In this work, we aim at studying $\rho-\omega$ mixing in a model
independent way by invoking Resonance Chiral
Theory(R$\chi$T)~\cite{Ecker}. It provides a reliable tool to study
physics in the intermediate energy
region~\cite{Guo:2011pa,Jamins,Roig:2013baa,chen2012,chen2014,Chen:2014yta}.
The tree-level calculation of $\rho-\omega$ mixing in the framework
of R$\chi$T has been given in Refs.~\cite{Urech,Ulf2006}, however,
the tree-level mixing amplitude turns out to be
momentum-independent. In order to implement the momentum dependence,
here we will calculate the one-loop contributions as shown in
Fig.~\ref{FeynmanDiagram}. The $\rho-\omega$ mixing can be induced
either by strong isospin-violating or by electromagnetic effects.
The former is proportional to the mass difference bewteen the $u,d$
quarks, i.e., $\Delta_{ud}=m_u-m_d$ and the latter is accompanied by
the fine structure constant $\alpha$. In the present study, only the
mixing effects linear in $\Delta_{ud}$ or $\alpha$ are under our
consideration. Apart from the overall factors $\Delta_{ud}$ or
$\alpha$, the large-$N_C$ counting rule proposed in
Ref.~\cite{Hooft} is imposed to truncate our perturbative
calculation. Specifically, our calculations are truncated at
next-to-leading order in the $1/N_C$ expansion for the strong and
electromagnetic contributions.  The ultraviolet (UV) divergence from
the loops is cancelled by introducing counterterms with sufficient
derivatives and the involved couplings are assumed to be beyond the
leading order in $1/N_C$ expansion as claimed in
Ref.~\cite{Rosell04}.

We assess the impact of momentum-dependent $\rho-\omega$ mixing
amplitude on the pion vector form factor by fitting to the
experimental data extracted from the $e^+e^-\rightarrow \pi^+\pi^-$
process and $\tau\rightarrow \nu_{\tau}2\pi$ decay in the energy
region of 650$\sim$850 MeV. Besides, the decay width of
$\omega\rightarrow \pi^+\pi^-$ is implemented as a constraint in the
fit. It is known that, provided isospin invariance holds, the
isovector part of the pion form factor in the $e^+e^-$ annihilation
is related to the one in $\tau$ decays theoretically, via the
conserved vector current assumption~\cite{CLEO,Davier03}. Different
effects of isospin breaking have been studied to describe the
$e^+e^-$ annihilation data and $\tau$ decays data simultaneously
\cite{Alemany,Oller:2000ug,Cirigliano01,Cirigliano02,Davier03,Ghozzi,Maltman,Dai:2013joa,Djukanovic:2014rua},
such as the short distance and long distance corrections in the
$\tau$ partial decay width to two pions, charged and neutral $\rho$
mass and width difference, and $\rho-\omega$ mixing. In our study we
will take into account all the above isospin breaking effects. Our
fit result shows that the $\rho-\omega$ mixing amplitude is
significantly momentum-dependent and its imaginary part is much
smaller than real part. Based on the fitted values of the
parameters, we also analyze the decay width of $\omega \rightarrow
\pi^+\pi^-$ by including the effect of the $\rho-\omega$ mixing. It
is found that the decay width is dominated by the $\rho-\omega$
mixing effect while the contribution from the direct coupling of
$\omega_I \rightarrow \pi^+\pi^-$ is negligible.

This paper is organized as follows. In Sec.~\ref{section.DescriptionOfMixing}, we introduce the description of
$\rho-\omega$ mixing. In Sec.~\ref{section.TheoreticalFramework}, we
present the theoretical framework and elaborate on the calculation
of the tree-level and loop contribution of $\rho-\omega$ mixing.
In Sec.~\ref{section.Phenomenology}, the fit result is shown and
the related phenomenology is dicussed. A summary is given in
Sec.~\ref{section.Conclusions}.

\section{Generic description of $\rho-\omega$
mixing}\label{section.DescriptionOfMixing}
In the isospin basis $|I,I_3\rangle$, we define $|\rho_I\rangle\equiv |1,0\rangle$ and $|\omega_I\rangle\equiv |0,0\rangle$ for convenience. The mixing between the isospin states of $|\rho_I\rangle$ and $|\omega_I\rangle$ can be implemented by considering the self-energy matrix
\bea \Pi_{\mu\nu}=T_{\mu\nu}\left(
                    \begin{array}{cc}
                      \Pi_{\rho\rho}(s) & \Pi_{\rho\omega}(s) \\
                      \Pi_{\rho\omega}(s) & \Pi_{\omega\omega}(s) \\
                    \end{array}
                  \right)\,,
\eea with $T_{\mu\nu}\equiv g_{\mu\nu}- \frac{p^\mu p^\nu}{p^2}$
\footnote{Without loss of generality, here we use the Proca
formalism for the vector fields and $T_{\mu\nu}$ is the transverse
projector. In the antisymmetric tensor formalism, the corresponding
transverse projector is
$\Omega^T_{\mu\nu\rho\sigma}=\frac{1}{2p^2}\big[(g_{\mu\rho} p_\nu
p_\sigma-g_{\rho\nu} p_\mu
p_\sigma)-(\rho\leftrightarrow\sigma)\big]$.  } and $s\equiv p^2$.
The off-diagonal matrix element $\Pi_{\rho\omega}(s)$ is none-zero,
e.g., due to isospin-breaking effect, and it therefore carries the
information of $\rho-\omega$ mixing. Subsequently, the dressed
propagator has the form~\cite{Connell97} \bea\label{eq:isoD}
D_{\mu\nu}=g_{\mu\nu}\left( \begin{array}{cc}
          {1}/{s_{\rho}} & \frac{\Pi_{\rho\omega}(s)}{s_{\rho}s_{\omega}} \\
          \frac{\Pi_{\rho\omega}(s)}{s_{\rho}s_{\omega}} & 1/s_{\omega}
         \end{array} \right)\equiv g_{\mu\nu}\, D^I(s)\ ,
         \eea
where the abbreviations $s_\rho$ and $s_\omega$ are defined by
 \bea\label{eq:srso} s_\rho&\equiv& s-\Pi_{\rho\rho}(s)-m_\rho^2,
\nonumber\\
s_\omega&\equiv& s-\Pi_{\omega\omega}(s)-m_\omega^2.
\eea
In above the vector-current conservation has been used to eliminate the longitudinal part proportional to $p_\mu$. Furthermore, we have also neglected terms of $\Pi_{\rho\omega}^2(s)$, since they correspond to contributions at two-loop order and are beyond our accuracy. $m_\rho$ and $m_\omega$ are bare masses of the $\rho$ and $\omega$ mesons, respectively.

The $\rho-\omega$ mixing, i.e., mixing between the physical states of $\rho^0$ and $\omega$, is obtainable by introducing  the following relation
\bea
  \left( \begin{array}{c}
          \rho^0 \\
          \omega
         \end{array} \right) =C \left( \begin{array}{c}
          \rho_I \\
          \omega_I
         \end{array} \right)\,,\qquad C= \left(\begin{array}{cc} 1 & -\epsilon_1 \\
\epsilon_2 & 1 \end{array} \right)\eea
with $\epsilon$ being the mixing parameter. The matrix of dressed propagators corresponding to physical states is diagonal.  Moreover, it can be connected to the matrix $D^I(s)$ in Eq.~(\ref{eq:isoD}) through
         \bea
  \left( \begin{array}{cc}
          1/s_{\rho} & 0 \\
          0 & 1/s_{\omega}
         \end{array} \right)=C
  \left( \begin{array}{cc}
          1/s_{\rho} & \Pi_{\rho\omega}/s_{\rho}s_{\omega} \\
          \Pi_{\rho\omega}/s_{\rho}s_{\omega} & 1/s_{\omega}
         \end{array} \right)
 C^{-1}. \eea
Solving the above equation and neglecting higher-order terms of
$\mathcal{O}(\epsilon^2)$ and $\epsilon\, \Pi_{\rho\omega}$,  one
obtains:
\begin{eqnarray} \epsilon_1=
\frac{\Pi_{\rho\omega}(M_\omega^2)}{s_\rho-s_\omega}.
\label{eqepsilon} \ ,\qquad \epsilon_2=
\frac{\Pi_{\rho\omega}(M_\rho^2)}{s_\rho-s_\omega}
\label{eqepsilon}\ .
\end{eqnarray}
The two mixing parameters should be just different with each other slightly, see Ref.~\cite{Connell97} for more details.

\section{Calculations in resonance chiral theory}
In this section we will calculate the mixing amplitude
$\Pi_{\rho\omega}(s)$ using R$\chi$T so as to study its momentum
dependence. The information of $\rho-\omega$ mixing is encoded in
the off-diagonal element of the self-energy matrix, which can be
decomposed as \bea
\Pi_{\rho\omega}(s)=\Delta_{ud}\,S_{\rho\omega}(s)+4\pi\alpha\,E_{\rho\omega}(s)\
, \eea where $\Delta_{ud}=m_u-m_d$ is the mass difference between
$u,d$ quarks, and $\alpha$ denotes the fine-structure constant. In
above, $S_{\rho\omega}(s)$ and $E_{\rho\omega}(s)$ stand for the
structure functions of strong and electromagnetic interactions,
respectively.  In the present work, the diagrams in
Fig.~\ref{FeynmanDiagram} are needed for a calculation in R$\chi$T
up to NLO in $1/N_C$ expansion. As will be seen below, the LO
contributions of $S_{\rho\omega}(s)$ and $E_{\rho\omega}(s)$ are
different: the former starts at $\mathcal{O}(N_C^0)$ while the
latter does at $\mathcal{O}(N_C^{1})$. Therefore, their
corresponding NLO contributions are of $\mathcal{O}(N_C^{-1})$ and
$\mathcal{O}(N_C^{0})$, respectively. In what follows, all the
diagrams in Fig.~\ref{FeynmanDiagram} will be calculated by using
effective Lagrangians constructed in the framework of R$\chi$T.

\begin{figure*}[ht]
\centering
\includegraphics[scale=0.6]{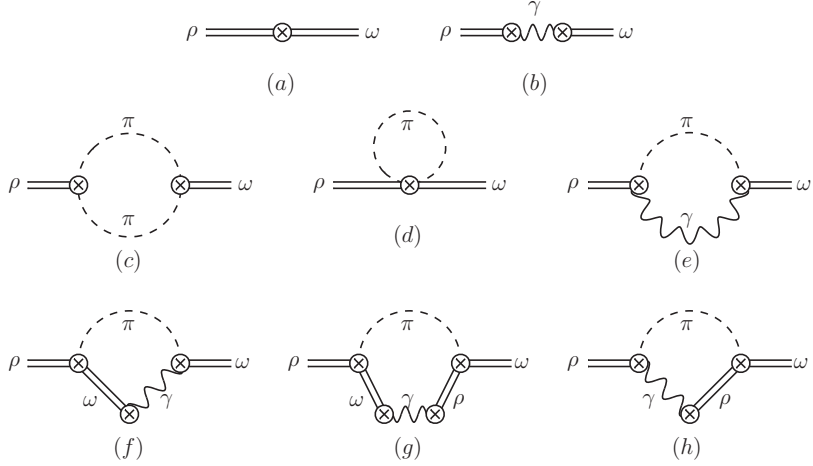}
\caption{ Feynman diagrams contributing to $\rho-\omega$
mixing. } \label{FeynmanDiagram}
\end{figure*}

\label{section.TheoreticalFramework}

\subsection{Resonance chiral theory and Tree-level amplitudes}

In R$\chi$T, the vector resonances are described in terms
of antisymmetry tensor fields with the normalization
\begin{eqnarray}
\langle 0|V_{\mu\nu}|V,p\rangle =iM_V^{-1}\{p_\mu\epsilon_\nu(p)-p_\nu\epsilon_\mu(p)\},
\label{eqnormalization}
\end{eqnarray} with $\epsilon_\mu$ being
the polarization vector. The kinetic Lagrangian of vector resonances takes the
form ~\cite{Ecker}
\begin{eqnarray}
\mathcal{L}_{kin}(V)=-\frac{1}{2}\langle \nabla^{\lambda}V_{\lambda\mu}\nabla_{\nu}V^{\nu\mu}-\frac{M_V^2}{2}V_{\mu\nu}V^{\mu\nu}\rangle \,,
\end{eqnarray}
where $M_V$ is the mass of the vector resonances in the chiral limit. Here the vector mesons are collected in a $2\times2$ matrix
\begin{equation}
V_{\mu\nu}=
 \left( \begin{array}{*{2}c}
   \frac{1}{\sqrt{2}}\rho^0 +\frac{1}{\sqrt{2}}\omega  & \rho^+    \\
   \rho^-  & -\frac{1}{\sqrt{2}}\rho ^0 +\frac{1}{\sqrt{2}}\omega   \\
\end{array} \right)_{\mu\nu}\,.
\end{equation}
Besides, the covariant derivative and chiral connection are defined by
\begin{eqnarray}
\nabla_{\mu}V&=&\partial_{\mu}V+[\Gamma_{\mu},V]\ ,\nonumber\\
\Gamma_{\mu}&=&\frac{1}{2}\{u^{+}(\partial _\mu-ir_\mu)u+u(\partial
_\mu-il_\mu)u^{+}\}.
\end{eqnarray}
The Goldstone bosons originating from the spontaneous breaking of the $SU(2)_L\times SU(2)_R$ chiral
symmetry are nonlinearly parametrized as
\begin{eqnarray}
u=\exp\{i\frac{\Phi}{\sqrt{2}F}\}\,,\qquad
\Phi=
 \left( {\begin{array}{*{2}c}
   {\frac{1}{\sqrt{2}}\pi ^0  } & {\pi^+ }   \\
   {\pi^- } & {-\frac{1}{\sqrt{2}}\pi ^0 }  \\
\end{array}} \right)\,,
\end{eqnarray}
with $F$ being the pion decay constant.

In the isospin limit, the standard Lagrangian describing the interactions between $V_{\mu\nu}$ and Goldstone bosons or electromagnetic fields are given by
\begin{eqnarray}\label{eq.interactionRchiT}
\mathcal{L}_{2}(V)&=&\frac{F_V}{2\sqrt{2}}\langle V_{\mu\nu}f_{+}^{\mu\nu}\rangle+\frac{iG_V}{\sqrt{2}}\langle V_{\mu\nu}u^\mu
u^\nu\rangle\ ,
\end{eqnarray}
with the relevant building blocks defined by
\begin{eqnarray}
f_{\pm}^{\mu\nu}&=&uF_L^{\mu\nu}u^{+}\pm u^{+}F_R^{\mu\nu}u\,, \nonumber\\
 u_\mu &=&i[u^{+}(\partial _\mu \
-ir_\mu)u-u(\partial _\mu -il_\mu)u^{+}] \,.
\end{eqnarray}
Here $F_{L,R}^{\mu\nu}$ are field strength tensors composed of the left and
right external sources $l_\mu$ and $r_\mu$, and $F_V, G_V$ are real
couplings.


The LO isospin-breaking effect is introduced by the Lagrangian
\begin{eqnarray}\label{eq.v8}
\mathcal{L}_2^{\rho\omega}=v_8\langle
V_{\mu\nu}V^{\mu\nu}\chi_+\rangle  \,,
\end{eqnarray}
with $\chi_+=u^{+}\chi u^{+}+ u\chi^{+}u$ and $\chi=2B_0(s+ip)$.
Here $v_8$ is an unknown coupling constant. However, it can be
determined by considering the mass relations of the vector mesons at
$O(p^2)$ in terms of the quark counting rule~\cite{Urech}, which
leads to: $v_8=1/8$.

With the above preparations, one is now able to calculate the tree amplitudes. The tree-level strong contribution, corresponding to diagram (a) in Fig.~\ref{FeynmanDiagram}, turns out to be
\begin{eqnarray}\label{eq.TreeAmplitude}
S_{\rho\omega}^{(a)}=2\,M_\rho \ ,
\end{eqnarray}
which is counted as $O(N_C^0)$, since $M_\rho\sim O(N_C^0)$. The
tree-level electromagnetic contribution is from diagram (b) in
Fig.~\ref{FeynmanDiagram} and {the amplitude can be obtained by
using the Lagrangian in Eq.~\eqref{eq.interactionRchiT}}:
\begin{eqnarray}\label{eq.TreeAmplitude}
E_{\rho\omega}^{(b)}=\frac{ F_\rho F_\omega}{3}\,,
\end{eqnarray}
with the $N_C$-counting order being $O(N_C^1)$ due to $F_\rho\sim
F_\omega\sim\sqrt{N_C}$. As mentioned in the beginning of this
section, the leading-order strong and electromagnetic contributions
indeed start at different orders in $1/N_C$ expansion.

\subsection{Loop contributions}
The relevant loop diagrams contributing up to our accuracy are shown in the second and third lines of Fig.~\ref{FeynmanDiagram}. Diagrams~(c) and~(d) contribute to the strong correction at $O(N_C^{-1})$, which are next-to-leading order compared to diagram~(a). Likewise, with respect to diagram~(b), diagrams~(e)-(h) lead to electromagnetic corrections at next-to-leading order, i.e. $O(N_C^0)$. In our calculation below, the necessary
isospin-breaking vertices are constructed based on the basic chiral
building blocks taken from $\chi$PT~\cite{gl845} and R$\chi$T~\cite{Ecker}.

\subsubsection{Diagram (c): $\pi\pi$ loop}

The vertex of $\rho_I\rightarrow \pi^+\pi^-$ can be read from the Lagrangian in
Eq.~\eqref{eq.interactionRchiT}. For the isospin-violating vertex of $\omega_I\rightarrow
\pi^+\pi^-$, we construct the following Lagrangian
\begin{eqnarray}
\mathcal{L}_{\omega_I\rightarrow
\pi^+\pi^-}&=&a_1i\langle V_{\mu\nu}\{\chi_+, u^\mu
u^\nu\}\rangle +a_2i\langle V_{\mu\nu}u^\mu \chi_+
u^\nu\rangle \nonumber\\
&=&(a_1-\frac{1}{2}a_2)\frac{8\sqrt{2}B_0i}{F^2}\Delta_{ud}\,\omega_{\alpha\beta}\pi^{+\alpha}\pi^{-\beta}.
\end{eqnarray}
For convenience, we define the combination $a\equiv a_1-\frac{1}{2}a_2.$
The $\pi\pi$-loop contribution can be obtained by calculating the integral
\begin{eqnarray}
i\Pi_{\rho\omega}\epsilon_{\rho\mu}\epsilon_\omega^\mu&=&-\frac{16\sqrt{2}G_VB_0a(m_u-m_d)p^2}{F^4}
\epsilon_{\rho\mu}\epsilon_{\omega\nu}\nonumber\\
&\times&\int
\frac{d^nk}{(2\pi)^n}\frac{k^\mu
k^\nu}{[k^2-m_\pi^2][(p-k)^2-m_\pi^2]}\ ,
\end{eqnarray}
where $p$ and $k$ denote the momenta of the external vector meson and either of the
exchanged pions, respectively. After integrating, the structure function can be extracted, which reads
 \begin{eqnarray}
S_{\rho\omega}^{(c)}
&=&
\frac{\sqrt{2}G_VB_0a}{12F^4\pi^2
}
p^4\bigg\{(1-\frac{6m_\pi^2}{p^2})(\lambda_{\infty}-\ln\frac{m_\pi^2}{\mu^2})\nonumber\\
&&\hspace{1.5cm}+\frac{5}{3}-\frac{8m_\pi^2}{p^2}
+\sigma^3\ln(\frac{\sigma+1}{\sigma-1})\bigg\},
\end{eqnarray}
where $\sigma\equiv\sqrt{1-4m_\pi^2/p^2}$
and $\lambda_{\infty}\equiv \frac{1}{\epsilon}-\gamma_E+1+\ln4\pi$ with
$\epsilon=2-\frac{d}{2}$ and $\gamma_E$ being the Euler constant.

\subsubsection{Diagram (d): $\pi$-tadpole loop}

According to the Lorentz, $P$ and $C$ invariances, the Lagrangian corresponding to the interaction of $\omega_I\rho_I \pi\pi$ can be written down as follows:
\begin{eqnarray}\label{eq.OmegaRhoPP}
\mathcal{L}_{\omega_I\rho_I PP}&=& b_1\langle V_{\mu\nu}V^{\mu\nu}(u^\alpha
u_\alpha \chi_+ +\chi_+u^\alpha u_\alpha)\rangle \nonumber\\
&+&b_2\langle V_{\mu\nu}V^{\mu\nu}u^\alpha \chi_+u_\alpha \rangle +b_3\langle V_{\mu\nu}\chi_+V^{\mu\nu}u^\alpha u_\alpha \rangle \nonumber \\
&+&
b_4\langle V_{\mu\nu}u^\alpha V^{\mu\nu} (\chi_+u_\alpha+u_\alpha\chi_+) \rangle \nonumber \\
&+&b_5\langle V_{\mu\alpha}V^{\nu\alpha}u^\mu u_\nu\chi_+ +V^{\nu\alpha}V_{\mu\alpha}\chi_+ u_\nu u^\mu\rangle \nonumber\\
&+&b_6\langle V_{\mu\alpha}V^{\nu\alpha}u^\mu\chi_+ u_\nu\rangle +b_7\langle V_{\mu\alpha}\chi_+V^{\nu\alpha}u^\mu u_\nu\rangle\nonumber \\
&+& b_8\langle V_{\mu\alpha}V^{\nu\alpha}u_\nu u^\mu \chi_+
+V^{\nu\alpha}V_{\mu\alpha}\chi_+  u^\mu u_\nu\rangle \nonumber \\
&+&b_9\langle V_{\mu\alpha}V^{\nu\alpha} u_\nu\chi_+u^\mu\rangle
+b_{10}\langle V_{\mu\alpha}\chi_+V^{\nu\alpha}u_\nu u^\mu \rangle \nonumber \\
&+&b_{11}\langle V_{\mu\alpha}u^\alpha V^{\mu\beta}u_\beta  \chi_+
+V^{\mu\beta}u^\alpha V_{\mu\alpha} \chi_+u_\beta \rangle \nonumber\\
&+&b_{12}\langle V_{\mu\alpha}u^\alpha V^{\mu\beta}\chi_+u_\beta
+V^{\mu\beta}u^\alpha V_{\mu\alpha} u_\beta\chi_+ \rangle \nonumber \\
&+&b_{13}\langle V_{\mu\alpha}u_\beta V^{\mu\beta}u^\alpha  \chi_+
+V^{\mu\beta}u_\beta V_{\mu\alpha} \chi_+u^\alpha \rangle\nonumber\\
&+&b_{14}\langle V_{\mu\alpha}u_\beta V^{\mu\beta}\chi_+u^\alpha
+V^{\mu\beta}u_\beta V_{\mu\alpha}u^\alpha \chi_+ \rangle \nonumber \\
&+&g_1i\langle V_{\mu\nu}V^{\mu\nu}(u^\alpha \nabla_\alpha\chi_-+
\nabla_\alpha\chi_-u^\alpha)\rangle \nonumber\\
&+&g_2i\langle V_{\mu\nu}u^\alpha V^{\mu\nu} \nabla_\alpha\chi_-\rangle \nonumber \\
&+&g_3i\langle V_{\mu\beta}V^{\mu\alpha}u^\beta \nabla_\alpha\chi_-+
V^{\mu\alpha}V_{\mu\beta}\nabla_\alpha\chi_-u^\beta \rangle \nonumber \\
&+&g_4i\langle V_{\mu\beta}V^{\mu\alpha} \nabla_\alpha\chi_-u^\beta+
V^{\mu\alpha}V_{\mu\beta}u^\beta\nabla_\alpha\chi_- \rangle \nonumber \\
&+&g_5i\langle V_{\mu\beta}u^\beta V^{\mu\alpha} \nabla_\alpha\chi_-+
V^{\mu\alpha}u^\beta V_{\mu\beta}\nabla_\alpha\chi_- \rangle \nonumber\\
&+&v_8\langle V_{\mu\nu}V^{\mu\nu}\chi_+\rangle \,.
\end{eqnarray}
Note that the $v_8\langle V_{\mu\nu}V^{\mu\nu}\chi_+\rangle $ term, which
contributes to the contact interaction of $\rho-\omega$ mixing, also
yields $\omega_I\rho_I \pi\pi$ vertex. Though in Eq.~\eqref{eq.OmegaRhoPP} there are many terms with a large number of free couplings, the final result only depends on combinations of these couplings. For simplicity, the following two combinations are necessary, i.e.,
\begin{eqnarray}
h_1&\equiv&6b_1-b_2+3b_3+b_4-2g_1-g_2, \nonumber\\
h_2&\equiv&4b_5-b_6+3b_7+4b_8-b_9+3b_{10}+2b_{11}+2b_{12}\nonumber\\
&&+2b_{13}+2b_{14}-2g_3-2g_4-2g_5\,.
\end{eqnarray}
Furthermore, one can neglect the mass difference between the charged and neutral pions in
the internal lines of loops, since the resultant difference is of higher orders beyond our consideration. As a result, the expanded form of Lagrangian~\eqref{eq.OmegaRhoPP} can be reduced simply to
\begin{eqnarray}
\mathcal{L}_{\omega_I\rho_I
\pi\pi}&=&\frac{4B_0}{F^2}h_1(m_u-m_d)\rho_{I\mu\nu}\omega^{\mu\nu}\pi_{\alpha}{\pi}^{\alpha}\nonumber\\
&-&\frac{2B_0}{F^2}v_8(m_u-m_d)\rho_{I\mu\nu}\omega^{\mu\nu}{\pi}^2\nonumber\\
&+&\frac{4B_0}{F^2}h_2(m_u-m_d)\rho_{I\mu\alpha}\omega^{\nu\alpha}\pi_{\mu}{\pi}^{\nu}\,.
\end{eqnarray}
With the above Lagrangian, the $\pi$-tadpole contribution to the $\rho-\omega$ mixing can be derived:
\begin{eqnarray}
i\Pi_{\rho\omega}\epsilon_{\rho\mu}\epsilon_\omega^\mu&=&
\frac{4iB_0}{F^2}h_1(m_u-m_d)\epsilon_{\rho\mu}\epsilon_\omega^\mu\int\frac{d^nk}{(2\pi)^n}
\frac{2ik^2}{k^2-m_\pi^2}\nonumber\\
&+&\frac{4iB_0}{F^2p^2}h_2(m_u-m_d)(p_\mu p_\nu
\epsilon_{\rho\alpha}\epsilon_\omega^\alpha+
p^2\epsilon_{\rho\mu}\epsilon_{\omega\mu})\nonumber\\
&\times&\int\frac{d^nk}{(2\pi)^n} \frac{ik^\mu k^\nu}{k^2-m_\pi^2}-i\frac{32B_0}{F^2}v_8(m_u-m_d)\nonumber\\
&\times&\epsilon_{\rho\mu}\epsilon_\omega^\mu\int\frac{d^nk}{(2\pi)^n}
\frac{i}{k^2-m_\pi^2}\ .
\end{eqnarray}
Eventually, the explicit expression of the strong structure function has the form of
\begin{eqnarray}\label{eq:tadpoleloop}
S^{(d)}_{\rho\omega}&
=&-\frac{m_\pi^2B_0}{8\pi^2F^2}\bigg\{(-16v_8+4h_1 m_\pi^2+h_2 m_\pi^2)\nonumber\\
&&\hspace{2cm}\times(\lambda_\infty-\ln\frac{m_\pi^2}{\mu^2})+\frac{h_2}{2}\bigg\}\,.
\end{eqnarray}

\subsubsection{Diagrams (e)-(h): $\pi^0\gamma$ loops}

In the loop diagrams (e)-(h), there are two types of vertices. The coupling of vector meson (V)  as well as vector external source (J) to pseudoscalar (P) is labeled by VJP vertex for short. The interaction of two vector mesons and one pseudoscalar is called VVP vertex. The operators of VJP type are given in
Ref.~\cite{Femenia}:
\begin{eqnarray} \label{eq.LagrangianVJP}
\mathcal{L}_{VJP}&=&\frac{c_1}{M_V}\epsilon_{\mu\nu\rho\sigma}\langle \{V^{\mu\nu},f_+^{\rho\alpha}\}\nabla_\alpha
u^\sigma\rangle \nonumber\\
&+&\frac{c_2}{M_V}\epsilon_{\mu\nu\rho\sigma}\langle \{V^{\mu\alpha},f_+^{\rho\sigma}\}\nabla_\alpha
u^\nu\rangle \nonumber\\
&+&\frac{ic_3}{M_V}\epsilon_{\mu\nu\rho\sigma}\langle \{V^{\mu\nu},f_+^{\rho\sigma}\}\chi_-\rangle \nonumber\\
&+&\frac{ic_4}{M_V}\epsilon_{\mu\nu\rho\sigma}\langle V^{\mu\nu}[f_-^{\rho\sigma},\chi_+]\rangle \nonumber\\
&+&\frac{c_5}{M_V}\epsilon_{\mu\nu\rho\sigma}\langle \{\nabla_\alpha
V^{\mu\nu},f_+^{\rho\alpha}\}u^\sigma\rangle \nonumber\\
&+&\frac{c_6}{M_V}\epsilon_{\mu\nu\rho\sigma}\langle \{\nabla_\alpha
V^{\mu\alpha},f_+^{\rho\sigma}\}u^\nu\rangle \nonumber\\
&+&\frac{c_7}{M_V}\epsilon_{\mu\nu\rho\sigma}\langle \{\nabla^\sigma
V^{\mu\nu},f_+^{\rho\alpha}\}u_\alpha\rangle \,,
\end{eqnarray}
and the ones of VVP type are
\begin{eqnarray}  \label{eq.LagrangianVVP}
\mathcal{L}_{VVP}&=&d_{1}\epsilon_{\mu\nu\rho\sigma}\langle \{V^{\mu\nu},V^{\rho\alpha}\}\nabla_\alpha
u^\sigma\rangle \nonumber\\
&+&id_{2}\epsilon_{\mu\nu\rho\sigma}\langle \{V^{\mu\nu},V^{\rho\sigma}\}\chi_-\rangle \nonumber\\
&+&d_{3}\epsilon_{\mu\nu\rho\sigma}\langle \{\nabla_\alpha
V^{\mu\nu},V^{\rho\alpha}\}u^\sigma\rangle \nonumber\\
&+&d_{4}\epsilon_{\mu\nu\rho\sigma}\langle \{\nabla^\sigma
V^{\mu\nu},V^{\rho\alpha}\}u_\alpha\rangle \,.
\end{eqnarray}
The involved couplings or their combinations can be estimated by
matching the leading operator product expansion of  $\langle
VVP\rangle $ Green function to the result calculted within R$\chi$T.
Such a procedure leads to high energy constraints on the resonance
couplings as follows~\cite{Femenia}:
\begin{eqnarray} \label{eq.HighEnergyConstraints}
4c_3+c_1&=&0,\nonumber\\ c_1-c_2+c_5&=&0,\nonumber\\
c_5-c_6&=&\frac{N_c}{64\pi^2}\frac{M_V}{\sqrt{2}F_V},\nonumber\\
d_1+8d_2&=&-\frac{N_c}{64\pi^2}\frac{M_V^2}{F_V^2}+\frac{F^2}{4F_V^2},\nonumber\\
d_3&=&-\frac{N_c}{64\pi^2}\frac{M_V^2}{F_V^2}+\frac{F^2}{8F_V^2}.
\end{eqnarray}
The mass of vectors in the chiral limit, $M_V$, can be estimated by
the mass of $\rho(770)$ meson~\cite{Mateu}.

The loops diagrams (e)-(h) can be calculated simultanously if the effective vertices of
$\rho^{\ast}\rightarrow\pi^{\ast}\gamma^{\ast}$ and
$\omega^{\ast}\rightarrow\pi^{\ast}\gamma^{\ast}$ are used, where a "$\ast$" stands for an off-shell particle. The explicit expression for $\rho^{\ast}\rightarrow\pi^{\ast}\gamma^{\ast}$ reads
\begin{eqnarray}
i\mathcal{V}^{\rho^{\ast}\pi^{\ast}\gamma^{\ast}}_{\rm eff}&=&i\epsilon_{\mu\nu\alpha\beta}k^\mu
p^\nu\epsilon_{\rho^{\ast}}^\alpha \epsilon_{\gamma^{\ast}}^\beta
\frac{4\sqrt{2}eB_0}{3M_\rho M_V F}[c_1(p-k)\cdot
k\nonumber\\
&-&c_2p\cdot(p-k)-4c_3m_\pi^2-c_5p\cdot k+c_6p^2]\nonumber\\
&+&i\epsilon_{\mu\nu\alpha\beta}k^\mu
p^\nu\epsilon_{\rho^{\ast}}^\alpha \epsilon_{\gamma^{\ast}}^\beta
\frac{4F_VeB_0}{3M_\rho F(M_\omega^2-k^2)}\nonumber\\
&\times&[d_1(p-k)^2
+8d_2m_\pi^2+2d_3p\cdot k]\ ,
\end{eqnarray}
where $p$ and $k$ denote the momentum of the vector meson and the
photon, respectively.
Analogically, for $\omega^{\ast}\rightarrow\pi^{\ast}\gamma^{\ast}$, one has
\begin{eqnarray}
i\mathcal{V}_{\rm eff}^{\omega^{\ast}\pi^{\ast}\gamma^{\ast}}&=&i\epsilon_{\mu\nu\alpha\beta}k^\mu
p^\nu\epsilon_{\omega^{\ast}}^\alpha \epsilon_{\gamma^{\ast}}^\beta
\frac{4\sqrt{2}eB_0}{M_\omega M_V F}[c_1(p-k)\cdot
k\nonumber\\
&-&c_2p\cdot(p-k)-4c_3m_\pi^2-c_5p\cdot k+c_6p^2]\nonumber\\
&+&i\epsilon_{\mu\nu\alpha\beta}k^\mu
p^\nu\epsilon_{\omega^{\ast}}^\alpha \epsilon_{\gamma^{\ast}}^\beta
\frac{4F_VeB_0}{M_\omega F(M_\rho^2-k^2)}\nonumber\\
&\times&[d_1(p-k)^2
+8d_2m_\pi^2+2d_3p\cdot k] \ .
\end{eqnarray}
It should be stressed that there are two terms in each effective vertex. One corresponds to the case that the virtual photon is coupled to the $VP$ system directly, while the other to  the case that it is interacted through an intermediate vector meson.
Note also that, throughout this work we only account for the corrections proportional either to $\Delta_{ud}$ or $4\pi\alpha$,  which implies the calculation of electromagnetic contribution can be carried out in the isospin limit, i.e., $m_u=m_d$.

With the help of the effective vertices, the $\pi\gamma$ loop contribution, i.e., the sum of the loops diagrams (e)-(h), can be expressed as:
\begin{eqnarray}
i\Pi_{\rho\omega}\epsilon_{\rho\mu}\epsilon_\omega^\mu&=&
\frac{1}{p^2}\int
\frac{d^nk}{(2\pi)^n}\frac{-i}{k^2}\frac{i}{(p-k)^2-m_\pi^2}\nonumber\\
&\times&[(k\cdot
p)^2\epsilon_{\rho}^\mu\epsilon_{\omega\mu}-k^2p^2\epsilon_{\rho}^\mu\epsilon_{\omega\mu}
+p^2k\cdot\epsilon_{\rho}k\cdot\epsilon_{\omega}]\nonumber\\
&\times&\bigg\{\frac{-32e^2}{3 M_V^2F^2}\big[c_1(p-k)\cdot
k-c_2(p-k)\cdot p\nonumber\\
&&\hspace{0.2cm}-4c_3m_\pi^2-c_5p\cdot k+c_6p^2\big]^2\nonumber\\
&-&\frac{16\sqrt{2}F_Ve^2}{3
M_VF^2}\bigg[\frac{1}{M_\omega^2-k^2}+\frac{1}{M_\rho^2-k^2}\bigg]\nonumber\\
&\times&\big[d_1(p-k)^2+8d_2m_\pi^2+2d_3p\cdot k\big]
\nonumber\\
&\times&
\big[c_1(p-k)\cdot k-c_2(p-k)\cdot p-4c_3m_\pi^2\nonumber\\
&-&c_5p\cdot k+c_6p^2\big]-\frac{16F_V^2e^2}{3
F^2(M_\rho^2-k^2)(M_\omega^2-k^2)}\nonumber\\
&\times&\big[d_1(p-k)^2+8d_2m_\pi^2+2d_3p\cdot
k\big]^2 \bigg\}\,.\hspace{10.5cm}
\end{eqnarray}
The further calculation is straightforward but the result of the extracted electromagnetic structure function $E^{\pi\gamma}_{\rho\omega}\equiv E_{\rho\omega}^{(e)}+E_{\rho\omega}^{(f)}+E_{\rho\omega}^{(g)}+E_{\rho\omega}^{(h)}$ is too lengthy to be shown here. It is worthy noting that in our numerical computation we will use the
high energy constraints in Eq.~\eqref{eq.HighEnergyConstraints} together with the fitted parameters given in
Ref.~\cite{chen2012}, therefore, all the parameters involved in $E^{\pi\gamma}_{\rho\omega}$ are known.

\subsubsection{ counterterms and renormalized amplitude }
Up to now, the total contribution of $\rho-\omega$ mixing can be expressed as
\bea
\Pi_{\rho\omega}&=&\Delta_{ud}\big[S_{\rho\omega}^{(a)}+S_{\rho\omega}^{(c)}+S_{\rho\omega}^{(d)}\big]+4\pi\alpha\big[E_{\rho\omega}^{(b)}+E_{\rho\omega}^{\pi\gamma}\big]\nonumber\ ,
\eea
which is still unrenormalized. The resonance chiral theory is unrenormalizable in the sense that the amplitude has to be renormalized order by order with increasing number of counterterms when the accuracy of the calculation is improved. In our case, the tree amplitudes, $S_{\rho\omega}^{(a)}$ and $E_{\rho\omega}^{(b)}$, can only absorb the ultraviolet divergence proportional to $p^0$. In order to cancel the $O(p^2)$, $O(p^4)$ and   $O(p^6)$ stemming from the loop contribution $S_{\rho\omega}^{(c)}$ and $E_{\rho\omega}^{\pi\gamma}$, additional counterterms are needed. For this purpose, we construct
\begin{eqnarray}
{\cal L}_{ct}&=&Y_A\langle V_{\mu\nu}V^{\mu\nu}\chi_+\rangle -\frac{1}{2}Y_B\langle \nabla^{\lambda}V_{\lambda\mu}\nabla_{\nu}V^{\nu\mu}\chi_+\rangle \nonumber\\
&+&\frac{Y_{C_1}}{2}\langle \nabla^2V^{\mu\nu}\{\chi_+,\{\nabla_\nu,\nabla^\sigma\}V_{\mu\sigma}\}\rangle \nonumber\\
&+&\frac{Y_{C_2}}{4}\langle \{\nabla_\nu,\nabla_\alpha\}V^{\mu\nu}\{\chi_+,\{\nabla^\sigma,\nabla^\alpha\}V_{\mu\sigma}\}\rangle \nonumber\\
&+&\frac{Y_{C_3}}{4}\langle \{\nabla^\sigma,\nabla^\alpha\}V^{\mu\nu}\{\chi_+,\{\nabla_\nu,\nabla_\alpha\}V_{\mu\sigma}\}\rangle \nonumber\\
&+&\frac{Z_A F_V}{2\sqrt{2}}\langle V_{\mu\nu}f_{+}^{\mu\nu}\rangle +\frac{Z_B
F_V}{2\sqrt{2}}\langle V_{\mu\nu}\nabla^{2}f_{+}^{\mu\nu}\rangle \nonumber\\
&+&\frac{Z_C
F_V}{2\sqrt{2}}\langle V_{\mu\nu}\nabla^{4}f_{+}^{\mu\nu}\rangle +\frac{Z_D
F_V}{2\sqrt{2}}\langle V_{\mu\nu}\nabla^{6}f_{+}^{\mu\nu}\rangle \,.
\end{eqnarray}
We  adopt the $\overline{\rm MS}-1$ subtraction scheme and
absorb the divergent pieces proportional to $\lambda_\infty$ by the bare couplings in the counterterms.  Consequently, the remanent finite pieces of counterterms can be written as:
\bea
\Pi_{\rho\omega}^{ct}=X_W^r\, p^6+X_Z^r\,p^4+X_R^r\, p^2 \ ,
\eea
with
\bea
X_W^r&\equiv& \frac{8\pi\alpha F_\rho F_\omega}{3}(Z_D^r +Z_B^r
Z_C^r)\ ,\nonumber\\
 X_Z^r&\equiv& \frac{4\pi\alpha F_\rho F_\omega}{3}(2 Z_C^r
+{Z_B^r}^2)\nonumber\\
&&+16M_\rho(m_u-m_d)(Y_{C_1}^r+Y_{C_2}^r+Y_{C_3}^r)\ ,\nonumber\\
X_R^r&\equiv&\frac{8\pi\alpha F_\rho
F_\omega}{3}Z_B^r-4M_\rho(m_u-m_d)Y_B^r\ . \eea In summary, the
UV-renormalized mixing amplitude reads
\begin{eqnarray} \label{eq.MixingAmplitude}
{\Pi}^r_{\rho\omega}(p^2)&=& 2M_\rho(m_u-m_d)+\frac{4\pi\alpha
F_\rho F_\omega}{3}\nonumber\\
&+&\frac{\sqrt{2}G_VB_0a}{12F^4\pi^2}
(m_u-m_d)p^4\bigg\{(\frac{6m_\pi^2}{p^2}-1)\ln\frac{m_\pi^2}{\mu^2}\nonumber\\
&+&\frac{5}{3}-\frac{8m_\pi^2}{p^2}
 +\sigma^3\ln(\frac{\sigma+1}{\sigma-1})\bigg\}\frac{m_\pi^2B_0}{8\pi^2F^2}(m_u-m_d)\nonumber\\
&\times&\bigg\{(-16v_8+4h_1m_\pi^2+h_2m_\pi^2)
\ln\frac{m_\pi^2}{\mu^2}-\frac{h_2}{2}\bigg\}\nonumber\\
&+&\overline{E}_{\rho\omega
}^{\pi\gamma}(p^2) +X_W^r p^6+ X_Z^rp^4+X_R^rp^2\,,
\end{eqnarray}
where a bar indicates the divergences are subtracted. As discussed in Ref.~\cite{Connell97}, there is  an important constraint on the mixing amplitude, namely, it should vanish as $p^2\to 0$. Thus the final expression of the renormalized mixing amplitude should be
\bea
{\Pi}_{\rho\omega}(p^2)={\Pi}^r_{\rho\omega}(p^2)-{\Pi}^r_{\rho\omega}(0)\ ,
\eea
where an additional finite shift is imposed so as to guarantee that the constraint ${\Pi}_{\rho\omega}(0)=0$ is satisfied.

In our numerical computation, the scale $\mu$ will be set to
$M_\rho$ and we use $(m_u-m_d)=-2.49$ MeV provided by particle data
group (PDG)~\cite{PDG2016}. Furthermore, we can define \bea
f_4&\equiv&\frac{m_\pi^2B_0}{8\pi^2F^2}(m_u-m_d)\big\{(-16v_8+4h_1m_\pi^2\nonumber\\
&&\qquad\qquad\quad+h_2m_\pi^2)
\ln\frac{m_\pi^2}{\mu^2}-\frac{h_2}{2}\big\}\ , \eea and in
principle the unknown parameters in Eq.~\eqref{eq.MixingAmplitude}
are $a$, $f_4$, $X_W^r$, $X_Z^r$ and $X_R^r$.

\begin{figure*}[t]
\centering
\includegraphics[height=9.5cm,width=18.5cm]{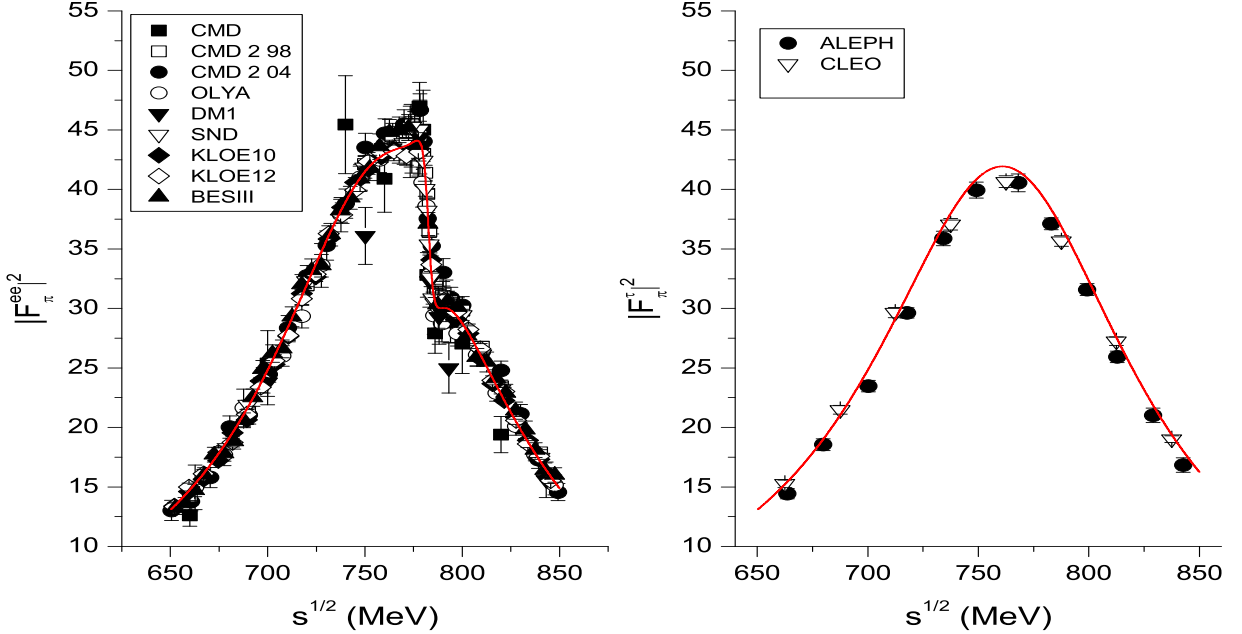}
\caption{  Fit results for the pion form factor in the $e^+e^-
\rightarrow \pi^+\pi^-$ process (left panel) and $\tau\rightarrow
\nu_\tau 2\pi$ process (right panel). The data of $e^+e^-$
annihilation are taken from the OLYA and CMD ~\cite{OLYACMD},
CMD2~\cite{CMD204,CMD2982}, DM1~\cite{DM1}, SND~\cite{SND},
KLOE~\cite{KLOE2011,KLOE2013}, BESIII~\cite{BESIII} collaborations.
The $\tau$ decay data are taken from the ALEPH~\cite{ALEPH} and
CLEO~\cite{CLEO} collaborations. The solid lines are our
theoretical predictions.} \label{fig2}
\end{figure*}

\section{The effect of  $\rho-\omega$ mixing on pion vector form factor} \label{section.Phenomenology}

The mass and width of $\rho$ meson  are conventionally determined by fitting to the data of $e^+e^-\rightarrow \pi^+\pi^-$ and
$\tau\rightarrow \nu_{\tau}2\pi$~\cite{PDG2016}, where various mechanisms are introduced to describe the $\rho-\omega$ mixing effect. To avoid intervening by their $\rho-\omega$ mixing mechanisms,  we do not employ their extracted values for the mass and width, rather,  we set the mass $M_\rho$, the relevant
couplings $G_\rho$ and $F_\rho$ to be free parameters in our fit. As for the width, a energy-dependent form will be imposed, which is supposed to be dominated by the two $\pi$ decay channel~\cite{Dumm}:
\begin{eqnarray}
\Gamma_{\rho}(s)=\frac{sM_\rho}{96\pi
F^2}(1-4m_\pi^2/s)^{\frac{3}{2}}.
\end{eqnarray}

For the narrow-width resonance $\omega$, we take
$M_\omega=782.65$ MeV and $\Gamma_\omega=8.49$ MeV from PDG
~\cite{PDG2016}. The physical coupling $F_\omega$ can be extracted from the
decay width of $\omega\rightarrow e^+e^-$. Using the Lagrangian
$\frac{F_V}{2\sqrt{2}}\langle V_{\mu\nu}f_{+}^{\mu\nu}\rangle $, one can derive the decay width
\begin{eqnarray}
\Gamma_{\omega}^{e^+e^-}=\frac{4\alpha^2\pi
F_\omega^2(2m_e^2+M_\omega^2)\sqrt{M_\omega^2-4m_e^2}}{27
M_\omega^4}\,,
\end{eqnarray}
and get $F_\omega\simeq 138$ MeV. With the decay widths given above, $s_\rho$ and $s_\omega$ in Eq.~\eqref{eq:srso} now can be rewritten as
\bea
s_\rho&\simeq& s-M_{\rho}^2+iM_{\rho}\Gamma_{\rho}(s) \ ,\nonumber\\
s_\omega&\simeq& s-M_{\omega}^2+iM_{\omega}\Gamma_{\omega}\ .
\eea

The experimental data considered in this work are the pion form
factor $F_\pi(p^2)$ of the $e^+e^-\rightarrow \pi^+\pi^-$
process~\cite{KLOE2011,KLOE2013,
BESIII,OLYACMD,CMD204,CMD2982,DM1,SND} and $\tau\rightarrow
\nu_{\tau}2\pi$ decay~\cite{ALEPH,CLEO} in the energy region of
650$\sim$850 MeV, and the decay width of $\omega \rightarrow
\pi^+\pi^-$~\cite{PDG2016}.

The Feynman amplitude for the process $\gamma^\ast
\rightarrow\pi^+\pi^-$, proceeding via virtual intermediate hadrons,
i.e., $\rho$, $\omega$ and their mixing, is described
by~\cite{Connell97}
\begin{eqnarray}
\mathscr{M}_{\gamma^\ast\rightarrow \pi\pi}
&=& \mathscr{M}_{\gamma^\ast\rightarrow\rho_I}\frac{1}{s_{\rho}}\mathscr{M}_{\rho_I\rightarrow\pi\pi}\nonumber\\
&+&\mathscr{M}_{\gamma^\ast\rightarrow\omega_I}\frac{1}{s_\omega}\Pi_{\rho\omega}\frac{1}{s_{\rho}}\mathscr{M}_{\rho_I\rightarrow\pi\pi}\nonumber\\
&+&\mathscr{M}_{\gamma^\ast\rightarrow\omega_I}\frac{1}{s_\omega}\mathscr{M}_{\omega_I\rightarrow\pi\pi}\nonumber\\
&+&\mathscr{M}_{\gamma^\ast\rightarrow\rho_I}\frac{1}{s_{\rho}}\Pi_{\rho\omega}\frac{1}{s_\omega}\mathscr{M}_{\omega_I\rightarrow\pi\pi}\,.
\nonumber
\end{eqnarray}

Here the fourth term corresponds to higher-order contribution of
isospin breaking, e.g., proportional to $(m_u-m_d)^2$, which is
beyond our accuracy and hence can be neglected. Including the
contribution from the direct coupling of photon to the pion pair,
the pion form-factor in $e^+e^-$ annihilation reads
\begin{eqnarray}
F_\pi^{ee}(p^2)&=&1-\frac{G_\rho F_\rho
p^2}{F^2}\frac{1}{s_{\rho}}-\frac{G_\rho F_\omega
p^2}{3F^2}\frac{1}{s_\omega}\Pi_{\rho\omega}\frac{1}{s_{\rho}}\nonumber\\
&-&\frac{4\sqrt{2}aB_0F_\omega(m_u-m_d)p^2}{3F^2}\frac{1}{s_\omega}\,,
\label{eqformfactoree}
\end{eqnarray}

On the other hand, the expression of the pion form-factor in $\tau\rightarrow \nu_{\tau}2\pi$ decay is
\begin{eqnarray}
F_\pi^\tau(p^2)=1-\frac{G_\rho F_\rho p^2}{F^2}\frac{1}{s_{\rho}}\,,
\end{eqnarray}
which is irrelevant to $\rho-\omega$ mixing effect. To take into account the
isospin breaking effects, one way is to multiply
$|F_\pi^\tau(p^2)|^2$ by the factor of $S_{EW}G_{EM}(s)$, where
$S_{EW}=1.0233$ corresponding to the short distance correction~\cite{Davier03}. Furthermore, $G_{EM}(s)$ is responsible for the long distance radiative correction  whose expression is provided in~\cite{Flores}. To be specific, in our fit we
make the following substitution
\bea |F_\pi^\tau(p^2)|^2\Rightarrow
S_{EW}G_{EM}(s)|F_\pi^\tau(p^2)|^2. \eea

Our best-fitted parameters and the corresponding
$\chi^2/\text{d.o.f.}$ are compiled in Table~\ref{table1}.  Our
determination of the mass of $\rho$ meson is in good agreement with the value reported in
PDG~\cite{PDG2016}. The fit
results are plotted in the Fig.~\ref{fig2}. One can see that the
experimental data of pion form factor, especially the kink around
the mass of $\omega$ in the $e^+ e^- \rightarrow \pi^+\pi^-$
process, is well described.

\begin{table}[h]
\begin{center}
\begin{tabular}{|c||c|}
\hline
         & Fit results  \\
\hline \hline
$M_{\rho}$ [MeV]     &    $ 775.3\pm0.3 $     \\
$G_\rho$ [MeV]     &    $ 67.0 \pm 3.0 $   \\
$F_\rho$ [MeV]     &    $152.9\pm 6.8 $    \\
$a~[\text{MeV}^{-1}]$         &    $(-1.8\pm 0.8)\times 10^{-6}$  \\
$X_W^r~[\text{MeV}^{-6}]$       &    $(7.3\pm 0.2)\times 10^{-17}$      \\
$X_Z^r~[\text{MeV}^{-4}]$         &   $(-5.5\pm 0.6)\times 10^{-11}$   \\
$X_R^r~[\text{MeV}^{-2}]$   &  $(-1.1\pm 0.1)\times 10^{-4}$     \\
$f_4^r~[\text{MeV}^{2}]$   &    $(1.3\pm 0.4)\times 10^{5}$  \\
\hline
$\chi^2/d.o.f$    & 314.9/(242-8)=1.35  \\
\hline
\end{tabular}
\caption{ The fit results of the parameters.}
\label{table1}
\end{center}
\end{table}

\begin{figure*}[t]
\centering
\includegraphics[scale=0.75]{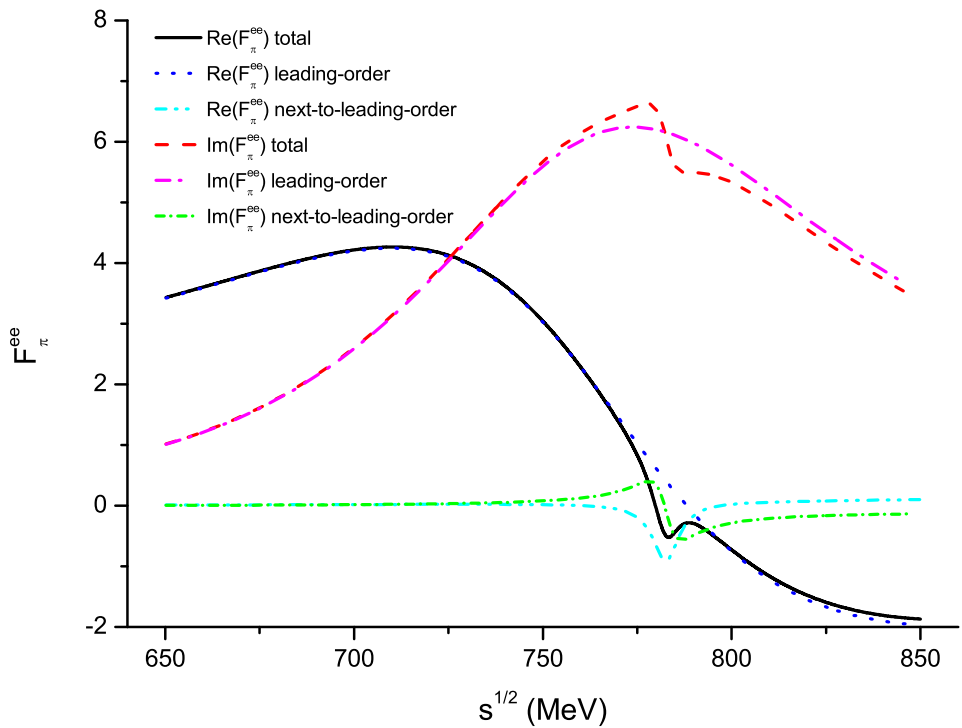}
\caption{ The real and imaginary parts of the fitted form factor
$F_\pi^{ee}(s)$. The black solid and red dashed lines represent our best results of the real and
imaginary parts, respectively. The blue dotted and cyan dash-dot-dotted lines correspond to the
leading order and the second order contributions of the real parts, respectively. The magenta dash-dotted
and green short dash-dash-dotted lines denote the leading order and second order contributions of the imaginary parts.} \label{fig3}
\end{figure*}

\begin{figure*}[t]
\centering
\includegraphics[height=9cm,width=18cm]{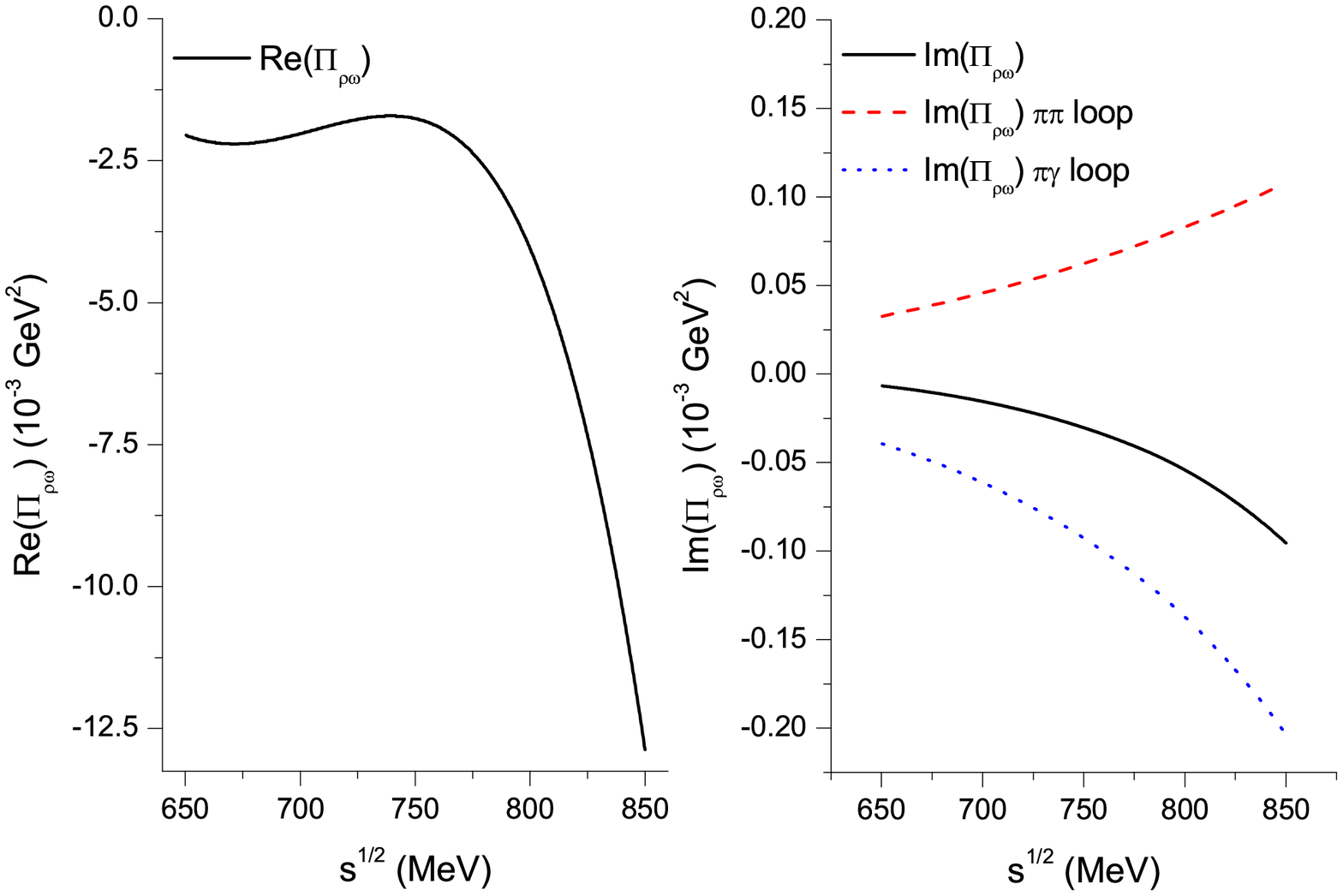}
\caption{ The real part (left panel) and imaginary part (right
panel) of the mixing amplitude
$\Pi_{\rho\omega}^{\text{physical}}(s)$. The black solid lines
represent our best fitted results. For the imaginary part, the red
dashed and blue dotted lines correspond to the contribution of
$\pi\pi$ loop and $\pi\gamma$ loop, respectively.} \label{fig4}
\end{figure*}

In Fig.~\ref{fig3},  contributions at different orders to the real
and imaginary parts of the pion form factor $F_\pi^{ee}(s)$ are
displayed. The leading-order contribution (mixing-effect irrelevant)
includes the contact interaction and the $\rho$-mediated mechanism,
namely the first two terms on the right side of
Eq.~\eqref{eqformfactoree}. The next-to-leading-order contribution
includes the $\rho-\omega$ mixing term and the direct $\omega_I
\pi\pi$ coupling, namely the third term plus the forth term on the
right side of Eq.~\eqref{eqformfactoree}. As expected, the
isospin-breaking effects mainly affects the
energy region around the masses of $\rho$ and $\omega$. It is found
that the dominant contribution is from the imaginary part in that
region. The isospin-breaking effects increase the absolute
value of imaginary part around the $\rho$ peak, and accounts for
that the $e^+e^-$ data are higher than the $\tau$ data in that
region as shown in Fig.~\ref{fig2}. Similar behavior has also been
observed in Ref.~\cite{Benayoun08} where the $\rho-\omega$ mixing
was treated in hidden local symmetry model.

In Fig.~\ref{fig4}, we plot the real and imaginary parts of the
mixing amplitude $\Pi_{\rho\omega}(s)$. It is found that the real
part is dominant almost in all the region and its
momentum-dependence is significant. Compared to the real part, the
imaginary part is rather small. For the imaginary part, the
contributions from $\pi\pi$ loop and $\pi\gamma$ loop are of the
same order, but with opposite sign. Note that the $\pi$-tadpole is
real and $s$-independent as can be seen from
Eq.~\eqref{eq:tadpoleloop}. The smallness of the imaginary part is
consistent with the observation in Refs.~\cite{Renard,Connell98},
though therein the effect of direct $\omega_I \rightarrow
\pi^+\pi^-$ was not taken into account and even in~\cite{Renard} the
isospin breaking is considered to be purely electromagnetic origin.
We also note that larger imaginary part is obtained
in~\cite{Mitchell,MLYan00} by using global color model and a chiral
constituent quark model, respectively. However, our finding is more
reliable in the sense that it is based on a model-independent
description of the $\rho-\omega$ mixing and, moreover, constraint
from experimental data is imposed by means of fitting.

The values of $\Pi_{\rho\omega}$ at physical masses of $\rho$ or
$\omega$ are interesting since they are related to the mixing parameters
given in Eq.~\eqref{eqepsilon}. To that end, we obtain: at
$s=M_\rho^2$, $\Pi_{\rho\omega}(M_\rho^2)=(-2380-40.8i)$ MeV$^2$ and
$\epsilon_2= 0.21$; at $s=M_\omega^2$,
$\Pi_{\rho\omega}(M_\omega^2)=(-2743.4-44.4i)$ MeV$^2$ and
$\epsilon_1=0.24$. As expected, $\epsilon_1$ and $\epsilon_2$ come out  to be almost the same.  Note that, in the numerical calculation of
$\epsilon_i$, we have neglected the small imaginary part of the mixing
amplitude as well as the widths of the $\rho$ and $\omega$
resonances. This leads to a real number of $\epsilon_i$ and hence a
probability interpretation can be assigned.

Using the central values of the
fitted parameters in Table~\ref{table1}, we calculate the decay width of
$\omega \rightarrow \pi^+\pi^-$
\begin{eqnarray}   \label{eq.Omegapipi}
\Gamma_{\omega \rightarrow \pi^+\pi^-}&=&\frac{1}{192\pi
F^4}(M_\omega^2-4m_\pi^2)^{\frac{3}{2}}\Big|
8\sqrt{2}B_0(m_u-m_d)a\nonumber\\
&&+\frac{2G_\rho
\Pi_{\rho\omega}(M_\omega^2)}{M_\omega^2-M_\rho^2-i(M_\omega
\Gamma_\omega-M_\rho \Gamma_\rho)}\Big|^2\nonumber\\
 &=&0.009\mid
(0.080)+(-0.446+3.783i)\mid^2 \,.
\end{eqnarray}
From Eq.~\eqref{eq.Omegapipi}, we can find that the first term due
to the direct $\omega_I \rightarrow \pi^+\pi^-$ is less than the
second term due to the $\rho-\omega$ mixing by two orders. In other words,
the direct $\omega_I\pi^+\pi^-$ coupling only affects the
decay width less than one percent. Within 1$\sigma$ uncertainties,
our theoretical value of the branching fraction is
$\mathscr{B}(\omega \rightarrow \pi^+\pi^-)=(1.53\pm 0.10)\times
10^{-2}$, which agrees with the values given in
PDG~\cite{PDG2016} and by the recent dispersive
analysis~\cite{Kubis2017}.

\section{Summary}  \label{section.Conclusions}

We have analyzed the $\rho-\omega$ mixing within the framework of
resonance chiral theory.  Based on the effective Lagrangians
constructed under the guidance of various symmetries, we calculate
the $\rho-\omega$ mixing amplitude up to next-to-leading order in
large $1/N_C$ expansion. Importantly, the momentum-dependent effect
is implemented due to the inclusion of loops in our calculation. The
values of the resonance couplings are determined by fitting to the
data of the pion vector form factor extracted from the
$e^+e^-\rightarrow \pi^+\pi^-$ process and $\tau\rightarrow
\nu_{\tau}2\pi$ decay. The decay width of $\omega\rightarrow
\pi^+\pi^-$ is served an additional constraint in the fit as well.
It is found that the imaginary part of the pion form factor
$F_\pi^{ee}(s)$ is enhanced largely around the $\rho$ peak. The
$\rho-\omega$ mixing amplitude is dominated by its real part almost
in all the region, which is significantly momentum-dependent. On the
contrary, its imaginary part is relatively small. We also find that
$\rho-\omega$ mixing plays a major role in the decay width of
$\omega \rightarrow \pi^+\pi^-$, and its contribution is two orders
of magnitude larger than that from the direct $\omega_I\pi\pi$
coupling.

\section*{Acknowledgments}
We would like to thank A.~Hosaka and J.~J.~Sanz-Cillero for helpful
discussions. This research is supported in part by the Fundamental
Research Funds for the Central Universities under Grant
No.~06500077, by the Spanish Ministerio de Econom\'ia y
Competitividad and the European Regional Development Fund, under
contracts FIS2014-51948-C2-1-P, FIS2014-51948-C2-2-P, SEV-2014-0398
and by Generalitat Valenciana under contract PROMETEOII/2014/0068.

\end{document}